\documentclass[reprint, amsmath, amssymb, aps]{revtex4-2}

\usepackage{graphicx}					
\usepackage{bm}							

\makeatletter							
\@ifclasswith{revtex4-2}{draft}{		
	\usepackage[inline]{showlabels}
	
}{}
\makeatother

\newcommand{\nn}{\nonumber}
\newcommand{\gl}{\big{(}}
\newcommand{\gr}{\big{)}}
\newcommand{\subt}[1]{_{\text{#1}}}

\newcommand{\qq}[1]{``#1''}

\newcommand{\eq}[1]{(\ref{eq:#1})}

\newcommand{\diag}{\mathrm{diag}}

\renewcommand{\i}{i}					
\renewcommand{\d}{\text{d}}
\newcommand{\vp}{\varphi}

\newcommand{\gtil}{\tilde{g}}
\newcommand{\cH}{\mathcal{H}}

\begin{document}

\title{\LARGE Physical time for the beginning universe}

\author{Christof Wetterich}
\affiliation{Institut f\"ur Theoretische Physik\\
Universit\"at Heidelberg\\
Philosophenweg 16, D-69120 Heidelberg}

\begin{abstract}

We propose that physical time is based on counting the oscillations of wave
functions. The discrete counting of the ticks of these clocks does not depend on
the metric frame. It remains well defined for the beginning epochs of the
universe. The photon clock counts the oscillations of electromagnetic plane
waves in the cosmic reference frame. It can be gauged with clocks for other
massless or massive particles. These equivalent clocks form a clock system which
defines a common universal continuous physical time. For the photon clock this
physical time coincides with conformal time. The universe is eternal towards the
past for typical models of inflationary cosmology -- the photon clock ticks an
infinite number of times. We compare the photon clock system to geodesic
physical time, which is a frame invariant generalization of proper time.

\end{abstract}

\maketitle



Time is a central concept for physics and much work has been devoted to define
\qq{physical time}, both theoretically and experimentally. When it comes to the
extreme situation in the beginning epochs of our universe the concept of
physical time may have to be rethought, however. There is no observer who can
construct some clock apparatus under the conditions of the radiation dominated
epoch or even earlier in cosmology. How should a physicist define time for these
extreme early epochs? The present contribution to this question is based on
numerous discussions I had with Valery Rubakov over the last two years of his
life. They resulted in the publication~\cite{RW}, expressing our common view,
with some points of our discussions left open.

In general relativity a physical time interval should not change under a general
coordinate transform. Physical time has to be diffeomorphism invariant. Most
commonly one uses for physical time the proper time $\tau$ measured by an
observer moving on some geodesics defined for particles with a fixed nonzero
mass $m$. An alternative could be conformal time $\eta$ which is associated to
massless particles as photons. The proper time depends on the geodesics chosen.
This is no problem as long as the proper times measured on two different
geodesics can be mapped onto each other by an invertible map. A given geodesic
defines a reference frame. The map between two inertial reference frames in flat
space is given by the Lorentz transformations of special relativity.

In cosmology one may choose a preferred \qq{cosmic reference frame} defined by
the approximate isotropy of the cosmic microwave background (CMB) in this frame.
For a spatially homogeneous and isotropic universe proper time in the cosmic
reference frame coincides with the time $t$ in the Robertson-Walker metric given
by
\begin{equation}
\label{eq:T1}
\d s^2=g_{\mu\nu}\d x^\mu\d x^\nu = -\d t^2 + a^2(t)\d x^i\d x^j\delta_{ij}\,.
\end{equation}
It is the proper time measured by a comoving observer whose trajectory obeys $\d
x^i = 0$. (The sign convention $(-,+,+,+)$ used in this note differs from
ref.~\cite{RW}.)

The notion of proper time measured on geodesics for massive particles has led to
the concept of \qq{geodesic completeness}. A geometry with Minkowski signature of
the metric is geodesically complete~\cite{PEN, HAW} if no geodesic stops at a
certain point. For a Robertson-Walker metric the criterion of geodesic
completeness is
\begin{equation}
\label{eq:T2}
\int_{-\infty}^\infty a(t)\d t = \infty\,.
\end{equation}
For a finite integral or $t$ not extending to $-\infty$ the geometry is
geodesically incomplete. A geodesically incomplete universe is often called
singular and it is believed that it cannot give a satisfactory description of the
beginning epoch. Typical models of inflationary cosmology have been shown to be
geodesically incomplete~\cite{BGV, MV}. This notion of geodesic completeness is
a purely geometric property which only depends on the form of the metric. It
does not involve any particle properties.

In many models for the beginning epochs of the universe scalar fields play an
important role, as the inflaton for inflationary cosmology. The dynamics of the
cosmological solution involves then an interplay between the scalar field and
the metric. In the presence of a dynamical scalar field pure geometric concepts
for physical time, as proper time for an observer moving on geodesics, have to
be questioned. In many models the masses of particles depend on the scalar
field, similar to the mass of the electron in the standard model which depends
on the value of the Higgs field. In cosmology the scalar field $\vp(x)$ is not
homogeneous in space and time -- for a dynamical scalar field it will depend at
least on time. In this case the particle masses $m(x)$ become space-time
dependent. This dependence induces an additional force (\qq{fifth force}) and
particles move no longer on geodesics. The trajectories $x^\mu(\sigma)$ obey now
a generalized geodesic equation~\cite{AWW, RW}
\begin{equation}
\label{eq:T3}
\frac{\d u^\mu}{\d\sigma} + \Gamma_{\nu\lambda}^\mu(x)u^\nu u^\lambda + \frac12
g^{\mu\nu}(x)\frac{\partial m^2(x)}{\partial x^\nu} = 0\,,
\end{equation}
where
\begin{equation}
\label{eq:T4}
u^\mu = \frac{\d x^\mu}{\d\sigma}\,,
\end{equation}
and $\sigma$ parameterizes the trajectory. Eq.~\eq{T3} is equivalent to
\begin{equation}
\label{eq:T5}
g_{\mu\nu}(x)u^\mu(\sigma)u^\nu(\sigma) = -m^2(x)\,.
\end{equation}
Only for constant $m$ eq.~\eq{T3} reduces to the geometric geodesic equation.
For general $m^2(x)$ particle physics properties influence the trajectories in
addition to the metric. This is easily understood by the presence of an
additional force. Evaluating proper time on the geometric geodesics makes not
much sense for particles (or observers) moving on different trajectories.

In the presence of a space-time dependent scalar field $\vp(x)$ the choice of
the metric is not unique. We could choose a different metric $\gtil_{\mu\nu}$
related to $g_{\mu\nu}$ by
\begin{equation}
\label{eq:T7}
\gtil_{\mu\nu}(x) = w^2\gl\vp(x)\gr g_{\mu\nu}(x)\,.
\end{equation}
The transformation from $g_{\mu\nu}$ to $\gtil_{\mu\nu}$ is a Weyl scaling or
conformal transformation. The choice of the metric is called a metric frame. If
one computes proper time from the metric $\gtil_{\mu\nu}$ it differs from the
one obtained from $g_{\mu\nu}$. Proper time is not a frame-invariant quantity.
Also the particle masses $m(x)$ are not frame invariant. The geometric geodesics
are obviously frame-dependent, and this extends to the notion of geodesic
completeness based on geometrical geodesics. Typical inflationary models are
geodesically incomplete in the Einstein frame for which the coefficient of the
curvature scalar in the effective action is given by the constant squared Planck
mass. These models can be mapped by a suitable choice of $w(\vp)$ to a so called
\qq{primordial flat frame} for which geometry becomes flat Minkowski space for
$t\to-\infty$~\cite{CWPFF}. In this frame the geometry is geodesically complete.
With geometric geodesic completeness depending on the metric frame one has to
rethink this concept.

In quantum field theory physical observables are independent of the choice of
fields. This is particularly apparent for the quantum effective action $\Gamma$
which includes all effects of quantum fluctuations. Similar to the classical
action $\Gamma$ is a functional of fields -- in this case the macroscopic or
mean fields. The quantum field equations obtain by variation of this effective
action with respect to the fields. For given $\Gamma$ these are exact equations
without any further corrections from quantum fluctuations. The correlation
function (connected two-point function) is given by the inverse second
functional derivative of $\Gamma$. Higher correlation functions (vertices) are
related to higher functional derivatives, such that the effective action
contains the full information about all observables. One can perform arbitrary
field transformations for the macroscopic fields which are the arguments of
$\Gamma$. Functional derivatives are transformed accordingly, as well as the
relations between observables and functional derivatives. At the end,
observables do not depend on the choice of the macroscopic fields -- see
ref.~\cite{CWFT} for a recent discussion. For cosmology, this basic property of
quantum field theory has been called \qq{field relativity}~\cite{CWFR}. It
states that the physics is independent of the choice of the metric frame.
Physics is not only invariant under general coordinate transformations, which
can be considered as a particular subclass of field transformations. It is
invariant under arbitrary field transformations, including the Weyl
transformations~\eq{T7}. General coordinate transformations leave the geometry
invariant, while this is no longer the case for more general field
transformations of the metric. In the view of field relativity geometry looses
its fundamental meaning.

These considerations have an important impact on the possible definitions of
physical time. Physical time should be a frame invariant quantity! As we have
seen this is not the case for proper time. We propose that the central concept
for physical time is the counting of oscillations of some periodic process. This
is actually the way how one defines the time standard on Earth -- one counts the
number of oscillations for the radiation emitted by some atomic transition. We
may define the periods by the zeros of some field component, e.g. the electric
field of the radiation. If these zeros are mapped to zeros in an invertible way
by a field transformation, the number of zeros remains the same. This is
realized by the frame transformation of the metric. Physical time defined by the
number of oscillations is a frame invariant quantity.

Oscillations occur in early cosmology. For example, we may consider the wave
function for the photon or the associated electromagnetic field in the cosmic
rest frame which obeys the field equation
\begin{equation}
\label{eq:T8}
(\partial_\eta^2+2\cH\partial_\eta+k^2)\vp(\vec{k}) = 0\,.
\end{equation}
Here $\eta = x^0$ is conformal time, $\d \eta=\d t/a$, $\cH=\partial_\eta\ln
a$, $g_{\mu\nu}=a^2(\eta)\eta_{\mu\nu}$, $\eta_{\mu\nu}=\diag(-1,1,1,1)$,
$k^2=\vec{k}^2$, and $\vec{k}$ is the wave vector (momentum) for a plane wave.
The field $\vp(\vec{k})$ stands for some component of the electromagnetic field
in Fourier space. It can also be a component of the graviton wave function or
any other massless field. An arbitrary electromagnetic field -- for example the
one of the CMB or the photons in the radiation dominated epoch -- can be
composed as a linear superposition of Fourier modes $\vp(\vec{k})$. For defining
a clock we focus on a particular wave number $\vec{k}$. For a given wave number
$\vec{k}$ the solution of the wave equation~\eq{T8} is a standing wave, which
reads in position space
\begin{equation}
\label{eq:T9}
\vp(\eta,\vec{x})=B(\eta)\exp\big\{\i(\vec{k}\vec{x} - k\eta)\big\}\,.
\end{equation}
The smooth function $B(\eta)$ accounts for the Hubble damping. For high momentum
modes with $k^2\gg\cH^2$ the function $B$ is essentially constant. For a given
position $\vec{x}$ one can count the number of oscillations $n_{\vec{k}}$.
Discrete physical time is proportional to $n_{\vec{k}}$, where the
proportionality coefficient sets the time units.

Plane waves with two different $\vec{k}$ constitute different clocks. The
number of oscillations is not the same, but proportional to each other. The
number of oscillations $n_{\vec{k_1}}$ of the first clock is related by an
invertible function to the number of oscillations $n_{\vec{k_2}}$ of the second
clock. The different clocks can therefore be gauged in order to define a unique
time by choosing appropriate proportionality coefficients. This is a first
example of a clock system. We call a clock system an ensemble of clocks for
which the ticks of any pair of two clocks can be mapped to each other by an
invertible function. It can be seen as an equivalence class of clocks. For the
clock system of the photon waves in the cosmic rest frame a useful continuous
physical time is conformal time $\eta$. For the high momentum modes the number
of oscillations is proportional to $\eta$
\begin{equation}
\label{eq:T10}
n_{\vec{k}} = \frac{k\eta}{2\pi}\,.
\end{equation}
Continuous time can be seen as the limit of discrete time for a very large
number of oscillations.

The clock system of photons can be extended to include oscillations for
gravitons or any other massless field. One may ask what happens for the
oscillations of the wave function for massive particles, or for oscillations in
different reference frames. Do they belong to the same clock system as the
\qq{photon clock}~\eq{T9}, or do they constitute different clock systems? For
late cosmology the map between all these different clocks is invertible. We can
express the number of ticks in terms of conformal time $\eta$. For example, the
ticks of an atomic clock on Earth can be expressed in terms of a corresponding
interval of conformal time. Similarly, proper time of an observer moving on
geodesics, or on more general physical trajectories, can be expressed in terms
of conformal time.

The issue of equivalence concerns the beginning of the universe. For the photon
clock system the physical time towards the beginning of the universe is infinite
for many models. For common inflationary models conformal time extends to
$-\infty$ as one moves backwards in cosmic time or some other time variable. The
number of ticks of the photon clocks is infinite since the \qq{beginning}. The
time-interval between two ticks shrinks if one uses cosmic time towards the
beginning, while it remains constant for conformal time. Still, the number of
ticks is the same in the two frames. If one uses the photon clock system for the
definition of physical time the universe exists since infinite time -- it is
eternal~\cite{CWEU}.

In contrast, for clocks defined by massive particles moving on suitable
trajectories the number of ticks may remain finite in the limit
$\eta\to-\infty$. In this case these clocks belong to a different clock system.
The map from such \qq{massive particle clocks} to the photon clocks does not
remain invertible for $\eta\to-\infty$. The issue of inequivalent clock systems
is directly related to the question: Does the universe exist since infinite
physical time?

In ref.~\cite{RW} we have studied the clocks for massive particles in detail, in
particular the case where the masses depend on some scalar field and therefore
on time. This leads to the concept of a frame-invariant generalization of proper
time for particles moving on generalized geodesics. We have found in some cases
clocks for massive particles that are not equivalent to the photon clocks. This
will bring us to the question which clocks are most useful for a definition of
physical time, which we will discuss at the end of this note. Our investigation
has suggested a frame-invariant generalization of the notion of geodesic
completeness, namely \qq{time completeness}. The universe is time-complete if
physical time is not bounded towards the beginning or end. The clocks defining
physical time never stop ticking, neither towards the beginning nor to the end.
If there exist inequivalent clock systems, time-completeness will depend on the
choice of the clock system. We briefly summarize here the results of
ref.~\cite{RW}.

Consider the clocks defined by a free scalar field with mass $m(x)$, minimal
coupling to gravity and canonical kinetic term. The field equation for a scalar
field with mass depending on the space-time coordinates $x$ reads
\begin{equation}
\label{eq:T11}
\gl g^{\mu\nu}(x)D_\mu\partial_\nu - m^2(x)\gr\vp(x) = 0\,,
\end{equation}
where $D_\mu$ is the covariant derivative in presence of the metric
$g_{\mu\nu}(x)$ and $g^{\mu\nu}(x)$ is the inverse metric. We look for fast
oscillations
\begin{equation}
\label{eq:T12}
\vp(x) = A(x)\exp\gl\i S(x)\gr\,,
\end{equation}
where $S(x)$ obeys
\begin{equation}
\label{eq:T13}
g^{\mu\nu}\partial_\mu S\partial_\nu S + m^2(x) = 0\,,
\end{equation}
and $A(x)$ is a slowly varying function. Since eq.~\eq{T13} is invariant under
Weyl scalings the quantity $S(x)$ is frame invariant. For counting oscillations
we have a tick of the clock whenever $S$ decreases by $2\pi$.

The definition of a clock needs a trajectory $x^\mu(\sigma)$ on which $S$ is
evaluated. In the late universe we may identify $x^\mu(\sigma)$ with the
trajectory of an observer. The oscillations of the scalar field are given by a
solution of eq.~\eq{T13} and do not depend on the presence of an observer, while
the number of zeros seen in some interval of $\sigma$ depend on the trajectory.
We can replace the observer by a particle moving on the trajectory
$x^\mu(\sigma)$. This definition of a \qq{particle clock} can be used for early
cosmology. A given trajectory defines a reference frame. For a given solution
for $S$ there exists a large family of clocks defined by different reference
frames. Our concept of physical time related to the oscillations of some field
induces in a natural way the notion of reference frames.

If cosmology is homogeneous and isotropic in the average one possible choice is
the \qq{cosmic reference frame}. It is given by a comoving trajectory with
constant $x^i$, $i=1\dots3$, and $x^0=c\sigma$. We could also use the physical
trajectories of particles with mass $\bar m(x)$ which obey eq.~\eq{T3}
defining the generalized geodesics. For $\bar m(x)=m(x)$ a particular clock
counts for a massive particle moving on a generalized geodesic the number of
oscillations of its own wave function. This \qq{geodesic physical time} is
analogous to proper time for an observer moving on geodesics. In contrast to
proper time the ticks of this clock are frame invariant.

We focus in the following on $\bar m(x) = m(x)$. In this case a family of
generalized geodesic trajectories can be related~\cite{RW} to the solution for
$S(x)$ by
\begin{equation}
\label{eq:T14}
\frac{\d x^\mu}{\d \sigma} = u^\mu = g^{\mu\nu}\frac{\partial S}{\partial
x^\nu}\,.
\end{equation}
From eqs.~\eq{T14},~\eq{T15} one concludes that $\d\sigma$ involves the
geometric proper time interval $\d\tau$,
\begin{equation}
\label{eq:T15}
\d\sigma = \frac{1}{m(x)}\sqrt{-g^{\mu\nu}(x)\d x^\mu\d x^\nu} =
\frac{\d\tau}{m(x)}\,.
\end{equation}
Neither $\d\sigma$ nor $\d\tau$ are frame invariant. If we compute the number of
ticks of the clock, however, we find a frame invariant quantity
\begin{align}
\label{eq:T16}
S(\sigma) - S(\sigma_0) = \int\d x^\mu\frac{\partial S}{\partial x^\mu} =& -\int
m^2(x)\d\sigma\nn\\
=& -\int m(x)\d\tau\,.
\end{align}
We define \qq{geodesic physical time} $T$ by
\begin{equation}
\label{eq:T17}
\d T = -\d S = m(x)\d\tau\,.
\end{equation}
The frame invariant generalization of the proper time interval $m(x)\d\tau$ has
been proposed in refs.~\cite{BST},~\cite{CWEU}. Since geodesic physical time is
invariant under Weyl transformations it can be evaluated in any metric frame
related by such transformations. One may introduce time units by multiplying the
dimensionless $\Delta t$ with a suitable proportionality constant. For constant
$m$ geodesic physical time coincides with proper time.

For a given solution $\vp(x)$ or $S(x)$ of the scalar field equation~\eq{T11}
or~\eq{T13} there are many trajectories which obey eq.~\eq{T14}. These
trajectories are distinguished by initial conditions $x^\mu(\sigma_0)$. For all
those trajectories eq.~\eq{T17} is valid. The clocks tick differently for
different trajectories since $m(x)$ and $\d\tau$ have to be evaluated for a
given trajectory. Furthermore, there are many trajectories which obey
the generalized geodesic equation~\eq{T3},~\eq{T4}, but do not obey
eq.~\eq{T14}. For those trajectories we can still count the periods and define
$\d T=-\d S$. The relation~\eq{T17} does not hold, however. All solutions of the
generalized geodesic equation define clocks in different reference frames. The
question arises how the clocks in different reference frames are related. For
flat space, constant $m$ and two trajectories specified by the four momenta
$p_\mu$ and $q_\mu$ of a massive particle one can show that the time intervals
$\Delta T$ in the two reference frames are related by a Lorentz
transformation~\cite{RW}. The general case with space-time dependent
$g_{\mu\nu}(x)$ and $m(x)$ can be rather complex. One finds that clocks defined
for different reference frames are not always equivalent -- they may belong to
different clock systems. There is no universal particle clock. Another
interesting issue concerns the question which particle clocks belong to the same
clock system as the photon clock.

We demonstrate these issues for a homogeneous isotropic cosmology with
Robertson-Walker metric~\eq{T1} and $m$ depending only on time. The wave
equation~\eq{T8} receives an additional contribution for massive particles
\begin{equation}
\label{eq:T18}
\gl\partial_\eta^2 + 2\cH\partial_\eta + k^2 + a^2m^2\gr\vp(\vec{k}) = 0\,.
\end{equation}
This results in
\begin{equation}
\label{eq:T19}
S = \tilde S(\eta) + k_ix^i\,,\quad \partial_\eta\tilde S = -\sqrt{k^2 +
a^2m^2}\,.
\end{equation}
For the rapid oscillations of interest both $a^2m^2$ and $\partial_\eta\cH$ are
small as compared to $k^2$. The solution of the field equation approaches the
one for massless particles.

Let us first consider the cosmic rest frame with trajectory
\begin{equation}
\label{eq:T20}
x^0 = \eta = c\sigma\,,\quad x^i = \text{const.}
\end{equation}
Physical time is given by
\begin{equation}
\label{eq:T21}
\Delta T = -\Delta S = -\int_{\eta\subt{in}}^\eta\d\eta\partial_\eta\tilde S =
\int_{\eta\subt{in}}^\eta\d\eta\sqrt{k^2+a^2m^2}\,.
\end{equation}
For all solutions with $k^2>0$ the physical time interval $\Delta T$ diverges
for $\eta\subt{in}\to-\infty$. These clocks are time complete. They belong to
the same clock system as the photon clock. In particular, if
$a^2(\eta)m^2(\eta)$ approaches zero for $\eta\subt{in}\to-\infty$, the ticks of
the particle clocks coincide in this limit with the ticks of the photon clock
for the same $k$.

We next consider the clocks for trajectories corresponding to the generalized
geodesics for the massive particle given by eq.~\eq{T14}
\begin{align}
\label{eq:T22}
u^0 =& -\frac{1}{a^2}\partial_\eta S = \frac{1}{a^2}\sqrt{k^2+m^2a^2}\,,\nn\\
u^i =& \phantom{-}\frac{1}{a^2}\partial_i S = \frac{1}{a^2}k^i\,.
\end{align}
One infers from eq.~\eq{T17} the geodesic physical time interval
\begin{equation}
\label{eq:T21A}
\d T = \frac{a^2m^2}{\sqrt{k^2+a^2m^2}}\d\eta\,.
\end{equation}
This differs from the result for the cosmic rest frame~\eq{T21}. If $a^2m^2$
remains finite for $\eta\to-\infty$ the geodesic physical time is equivalent to
the one defined by the photon clock. On the other hand, if $a^2m^2\to0$ for
$\eta\subt{in}\to-\infty$, one has
\begin{equation}
\label{eq:T24}
\Delta T = \frac{1}{k}\int_{\eta\subt{in}}^\eta\d\eta a^2(\eta)m^2(\eta)\,.
\end{equation}
For
\begin{equation}
\label{eq:T25}
am = |\eta|^{-\alpha}\,,
\end{equation}
the geodesical physical time interval remains finite for
$\eta\subt{in}\to-\infty$ if $\alpha\geq1/2$. In this case the particle clocks
on trajectories given by generalized geodesics~\eq{T14} do not belong to the
same clock system as the photon clock. They are not time complete or
\qq{geodesically complete} in a generalized sense. For example, this happens for
de Sitter space with constant $m$ where $\alpha = 1$. Since $\d T$ is frame
invariant, the same happens in the corresponding primordial flat frame where $a$
approaches a constant for $\eta\to-\infty$, while $m$ vanishes in this limit.

One would like to understand why the particle clocks on generalized geodesics
can stop ticking for $\eta\to-\infty$, whereas the photon clocks tick at a
constant pace. This is best understood in a frame where $a$ is constant, while
$m$ vanishes for $\eta\to-\infty$. With $a=1$ conformal time equals cosmic time
and the photon wave functions are simple plane waves. In lowest order in $m/k$
the particle wave function obeys
\begin{equation}
\label{eq:T26}
S = \vec{k}\vec{x} - k\eta - \frac{1}{2k}\int_{-\infty}^\eta\d\eta'm^2(\eta')\,.
\end{equation}
The wave front $S = 0$, $\vec{x} = x\subt{w}\vec{k}/k$, obeys
\begin{equation}
\label{eq:T27}
x\subt{w} = \eta + \frac{1}{2k^2}\int_{-\infty}^\eta\d\eta' m^2(\eta')\,,\quad
\partial_\eta x\subt{w} = 1+\frac{m^2}{2k^2}\,.
\end{equation}
The trajectory of the particle moving on generalized geodesics,
$\vec{x}=x\subt{p}\vec{k}/k$, can be characterized by its velocity,
\begin{equation}
\label{eq:T28}
\partial_\eta x\subt{p} = \frac{k}{\sqrt{k^2+m^2}} \approx 1-\frac{m^2}{2k^2}\,.
\end{equation}
For $m>0$ the particle is slower than the wave front, $\partial_\eta
x\subt{p}<\partial_\eta x\subt{w}$. In the reference frame of the particle one
can count the oscillations of the wave function which overtakes it. These
oscillations determine the ticks of the clock in the particle rest frame. In the
limit $m\to0$ the particle velocity and the wave front velocity approach each
other. From the point of view of the particle rest frame the wave front gets
frozen. The interval between two ticks gets longer and longer. If $m$ decreases
fast enough, the geodesic particle clocks stops ticking for $\eta\to-\infty$.

In the presence of two inequivalent clock systems the debate is open which clock
system is best suited for a description of physical reality. The discussion on
this issue was ongoing between Valery Rubakov and me~\cite{RW}. An advantage of
geodesical physical time for particle clocks is its formulation for arbitrary
metrics and arbitrary trajectories. On the other side, one may consider as an
advantage of the photon clock system that waves of massless particles as photons
or gravitons exist for the beginning epoch of all realistic cosmologies. For
the cosmic reference frame the limit of a particle getting massless is smooth.

An open point for the photon clock system is the formulation of a generalized
cosmic rest frame for arbitrary inhomogeneous metrics. Oscillating wave
functions for photons or gravitons exist for rather arbitrary metrics, at least
if deviations from homogeneity and isotropy remain small. The issue concerns
the choice of reference frames. For weak inhomogeneities of the metric there are
large families of trajectories for which the photon clock is time complete. As
long as inhomogeneities are moderate one may define an isotropic homogeneous
rest frame by a suitable averaging of the metric. Further work is necessary in
order to clarify this issue. For most practical applications in cosmology one
can simply identify physical time for the photon clock system with conformal
time for the \qq{background metric}. For typical inflationary cosmologies the
universe is eternal.

In conclusion, the definition of physical time by counting oscillations of wave
functions is well adapted to the beginning epoch of the universe when no
observers with a measuring apparatus have existed. The discrete counting does not
depend on the metric frame, in contrast to proper time. For a given solution of
the wave function the number of oscillations, which corresponds to the number of
ticks of a clock, depends on the trajectory on which the counting is done. This
introduces the notion of reference frames in a wide context.

A rather natural clock is the photon clock which counts the oscillations of
electromagnetic plane waves in the cosmic reference frame. For the cosmic
reference frame all clocks defined by the oscillations of other massless fields
as the metric, or equivalently of the wave functions for massive particles,
belong to the same clock system as the photon clock. They can be mutually gauged
to each other. The corresponding universal time can be identified with conformal
time. Clocks in the cosmic reference frame tick an infinite number of times as
one moves backward for most common inflationary models. For these models the
universe is time complete or eternal.

Geodesic physical time counts for a massive particle moving on a physical
trajectory the number of oscillations of its own wave function. The physical
trajectories are generalized geodesics which take into account additional forces
for a space-time dependent mass. Geodesic physical time generalizes proper time
for the case of varying mass. In contrast to proper time it is a frame-invariant
quantity. For certain cases where the particle mass or the scale factor vanish
too rapidly towards the beginning of the universe the geodesic particle clock
stops ticking in this limit. This reflects the well known fact that the rest
frame is not well defined for massless particles.

The geodesic physical time can be defined for arbitrary inhomogeneous metrics
and particle masses. For the photon clock further work is needed for the
generalization of the cosmic rest frame to arbitrary inhomogeneous metrics.
Since massless particles, or effectively massless particles are omnipresent in
models of the beginning universe, the photon clock system seems to me to be the
most useful definition of physical time. It can be used even if the beginning is
characterized by quantum scale symmetry where no intrinsic mass or length scale
exists and all excitations or fluctuations are massless.


\begin{acknowledgments}
This note is dedicated to my friend Valery, whose deep physical insights and
profound honesty and engagement I have admired.
\end{acknowledgments}




\nocite{*}
\bibliography{refs}

\begin{thebibliography}{11}%
\makeatletter
\providecommand \@ifxundefined [1]{%
 \@ifx{#1\undefined}
}%
\providecommand \@ifnum [1]{%
 \ifnum #1\expandafter \@firstoftwo
 \else \expandafter \@secondoftwo
 \fi
}%
\providecommand \@ifx [1]{%
 \ifx #1\expandafter \@firstoftwo
 \else \expandafter \@secondoftwo
 \fi
}%
\providecommand \natexlab [1]{#1}%
\providecommand \enquote  [1]{``#1''}%
\providecommand \bibnamefont  [1]{#1}%
\providecommand \bibfnamefont [1]{#1}%
\providecommand \citenamefont [1]{#1}%
\providecommand \href@noop [0]{\@secondoftwo}%
\providecommand \href [0]{\begingroup \@sanitize@url \@href}%
\providecommand \@href[1]{\@@startlink{#1}\@@href}%
\providecommand \@@href[1]{\endgroup#1\@@endlink}%
\providecommand \@sanitize@url [0]{\catcode `\\12\catcode `\$12\catcode
  `\&12\catcode `\#12\catcode `\^12\catcode `\_12\catcode `\%12\relax}%
\providecommand \@@startlink[1]{}%
\providecommand \@@endlink[0]{}%
\providecommand \url  [0]{\begingroup\@sanitize@url \@url }%
\providecommand \@url [1]{\endgroup\@href {#1}{\urlprefix }}%
\providecommand \urlprefix  [0]{URL }%
\providecommand \Eprint [0]{\href }%
\providecommand \doibase [0]{https://doi.org/}%
\providecommand \selectlanguage [0]{\@gobble}%
\providecommand \bibinfo  [0]{\@secondoftwo}%
\providecommand \bibfield  [0]{\@secondoftwo}%
\providecommand \translation [1]{[#1]}%
\providecommand \BibitemOpen [0]{}%
\providecommand \bibitemStop [0]{}%
\providecommand \bibitemNoStop [0]{.\EOS\space}%
\providecommand \EOS [0]{\spacefactor3000\relax}%
\providecommand \BibitemShut  [1]{\csname bibitem#1\endcsname}%
\let\auto@bib@innerbib\@empty
\bibitem [{\citenamefont {Rubakov}\ and\ \citenamefont {Wetterich}(2022)}]{RW}%
  \BibitemOpen
  \bibfield  {author} {\bibinfo {author} {\bibfnamefont {V.~A.}\ \bibnamefont
  {Rubakov}}\ and\ \bibinfo {author} {\bibfnamefont {C.}~\bibnamefont
  {Wetterich}},\ }\href@noop {} {\bibinfo {title} {Geodesic (in)completeness in
  general metric frames}} (\bibinfo {year} {2022}),\ \Eprint
  {https://arxiv.org/abs/2210.11198} {arXiv:2210.11198 [gr-qc]} \BibitemShut
  {NoStop}%
\bibitem [{\citenamefont {Penrose}(1965)}]{PEN}%
  \BibitemOpen
  \bibfield  {author} {\bibinfo {author} {\bibfnamefont {R.}~\bibnamefont
  {Penrose}},\ }\bibfield  {title} {\bibinfo {title} {Gravitational collapse
  and space-time singularities},\ }\href@noop {} {\bibfield  {journal}
  {\bibinfo  {journal} {Phys. Rev. Lett.}\ }\textbf {\bibinfo {volume} {14}},\
  \bibinfo {pages} {57} (\bibinfo {year} {1965})}\BibitemShut {NoStop}%
\bibitem [{\citenamefont {Hawking}(1966)}]{HAW}%
  \BibitemOpen
  \bibfield  {author} {\bibinfo {author} {\bibfnamefont {S.~W.}\ \bibnamefont
  {Hawking}},\ }\bibfield  {title} {\bibinfo {title} {Singularities in the
  universe},\ }\href@noop {} {\bibfield  {journal} {\bibinfo  {journal} {Phys.
  Rev. Lett.}\ }\textbf {\bibinfo {volume} {17}},\ \bibinfo {pages} {444}
  (\bibinfo {year} {1966})}\BibitemShut {NoStop}%
\bibitem [{\citenamefont {Borde}\ \emph {et~al.}(2003)\citenamefont {Borde},
  \citenamefont {Guth},\ and\ \citenamefont {Vilenkin}}]{BGV}%
  \BibitemOpen
  \bibfield  {author} {\bibinfo {author} {\bibfnamefont {A.}~\bibnamefont
  {Borde}}, \bibinfo {author} {\bibfnamefont {A.~H.}\ \bibnamefont {Guth}},\
  and\ \bibinfo {author} {\bibfnamefont {A.}~\bibnamefont {Vilenkin}},\
  }\bibfield  {title} {\bibinfo {title} {Inflationary spacetimes are incomplete
  in past directions},\ }\href@noop {} {\bibfield  {journal} {\bibinfo
  {journal} {Physical Review Letters}\ }\textbf {\bibinfo {volume} {90}}
  (\bibinfo {year} {2003})}\BibitemShut {NoStop}%
\bibitem [{\citenamefont {Mithani}\ and\ \citenamefont {Vilenkin}(2012)}]{MV}%
  \BibitemOpen
  \bibfield  {author} {\bibinfo {author} {\bibfnamefont {A.}~\bibnamefont
  {Mithani}}\ and\ \bibinfo {author} {\bibfnamefont {A.}~\bibnamefont
  {Vilenkin}},\ }\href@noop {} {\bibinfo {title} {Did the universe have a
  beginning?}} (\bibinfo {year} {2012}),\ \Eprint
  {https://arxiv.org/abs/1204.4658} {arXiv:1204.4658 [hep-th]} \BibitemShut
  {NoStop}%
\bibitem [{\citenamefont {Ayaita}\ \emph {et~al.}(2012)\citenamefont {Ayaita},
  \citenamefont {Weber},\ and\ \citenamefont {Wetterich}}]{AWW}%
  \BibitemOpen
  \bibfield  {author} {\bibinfo {author} {\bibfnamefont {Y.}~\bibnamefont
  {Ayaita}}, \bibinfo {author} {\bibfnamefont {M.}~\bibnamefont {Weber}},\ and\
  \bibinfo {author} {\bibfnamefont {C.}~\bibnamefont {Wetterich}},\ }\bibfield
  {title} {\bibinfo {title} {Structure formation and backreaction in growing
  neutrino quintessence},\ }\href@noop {} {\bibfield  {journal} {\bibinfo
  {journal} {Physical Review D}\ }\textbf {\bibinfo {volume} {85}} (\bibinfo
  {year} {2012})},\ \Eprint {https://arxiv.org/abs/1112.4762} {arXiv:1112.4762
  [astro-ph.CO]} \BibitemShut {NoStop}%
\bibitem [{\citenamefont {Wetterich}(2021)}]{CWPFF}%
  \BibitemOpen
  \bibfield  {author} {\bibinfo {author} {\bibfnamefont {C.}~\bibnamefont
  {Wetterich}},\ }\bibfield  {title} {\bibinfo {title} {Primordial flat frame:
  A new view on inflation},\ }\href@noop {} {\bibfield  {journal} {\bibinfo
  {journal} {Physical Review D}\ }\textbf {\bibinfo {volume} {104}} (\bibinfo
  {year} {2021})}\BibitemShut {NoStop}%
\bibitem [{\citenamefont {Wetterich}(2024)}]{CWFT}%
  \BibitemOpen
  \bibfield  {author} {\bibinfo {author} {\bibfnamefont {C.}~\bibnamefont
  {Wetterich}},\ }\href@noop {} {\bibinfo {title} {Field transformations in
  functional integral, effective action and functional flow equations}}
  (\bibinfo {year} {2024}),\ \Eprint {https://arxiv.org/abs/2402.04679}
  {arXiv:2402.04679 [hep-th]} \BibitemShut {NoStop}%
\bibitem [{\citenamefont {Wetterich}(2013)}]{CWFR}%
  \BibitemOpen
  \bibfield  {author} {\bibinfo {author} {\bibfnamefont {C.}~\bibnamefont
  {Wetterich}},\ }\bibfield  {title} {\bibinfo {title} {Universe without
  expansion},\ }\href@noop {} {\bibfield  {journal} {\bibinfo  {journal}
  {Physics of the Dark Universe}\ }\textbf {\bibinfo {volume} {2}},\ \bibinfo
  {pages} {184–187} (\bibinfo {year} {2013})},\ \Eprint
  {https://arxiv.org/abs/1303.6878} {arXiv:1303.6878 [astro-ph.CO]}
  \BibitemShut {NoStop}%
\bibitem [{\citenamefont {Wetterich}(2014)}]{CWEU}%
  \BibitemOpen
  \bibfield  {author} {\bibinfo {author} {\bibfnamefont {C.}~\bibnamefont
  {Wetterich}},\ }\bibfield  {title} {\bibinfo {title} {Eternal universe},\
  }\href@noop {} {\bibfield  {journal} {\bibinfo  {journal} {Physical Review
  D}\ }\textbf {\bibinfo {volume} {90}} (\bibinfo {year} {2014})},\ \Eprint
  {https://arxiv.org/abs/1404.0535} {arXiv:1404.0535 [gr-qc]} \BibitemShut
  {NoStop}%
\bibitem [{\citenamefont {Bars}\ \emph {et~al.}(2013)\citenamefont {Bars},
  \citenamefont {Steinhardt},\ and\ \citenamefont {Turok}}]{BST}%
  \BibitemOpen
  \bibfield  {author} {\bibinfo {author} {\bibfnamefont {I.}~\bibnamefont
  {Bars}}, \bibinfo {author} {\bibfnamefont {P.~J.}\ \bibnamefont
  {Steinhardt}},\ and\ \bibinfo {author} {\bibfnamefont {N.}~\bibnamefont
  {Turok}},\ }\bibfield  {title} {\bibinfo {title} {Cyclic cosmology, conformal
  symmetry and the metastability of the higgs},\ }\href@noop {} {\bibfield
  {journal} {\bibinfo  {journal} {Physics Letters B}\ }\textbf {\bibinfo
  {volume} {726}},\ \bibinfo {pages} {50–55} (\bibinfo {year} {2013})},\
  \Eprint {https://arxiv.org/abs/1303.6878} {arXiv:1303.6878 [gr-qc]}
  \BibitemShut {NoStop}%
\end{thebibliography}%

\end{document}